\documentclass{IEEEcsmag}

\usepackage[colorlinks,urlcolor=blue,linkcolor=blue,citecolor=blue]{hyperref}

\usepackage{upmath}

\usepackage{epsfig}
\usepackage{graphicx}
\usepackage{subfigure}
\usepackage{wrapfig}
\usepackage{amsmath, amsfonts, amsthm, amssymb}
\usepackage{algorithm}
\usepackage{algorithmic}
\usepackage{alltt}
\usepackage{url}
\usepackage{color,soul}

\usepackage{xcolor}
\usepackage{tcolorbox}

\begin{document}

\title{Confidential High-Performance Computing in the Public Cloud}

\author{Keke Chen}
\affil{Computer Science Department, Marquette University}

\begin{abstract}
High-Performance Computing (HPC) in the public cloud democratizes the supercomputing power that most users cannot afford to purchase and maintain. Researchers have studied its viability, performance, and usability. However, HPC in the cloud has a unique feature -- users have to export data and computation to somewhat untrusted cloud platforms. Users will either fully trust cloud providers to protect from all kinds of attacks or keep sensitive assets in-house instead. With the recent deployment of the Trusted Execution Environment (TEE) in the cloud, confidential computing for HPC in the cloud is becoming practical for addressing users' privacy concerns. This paper discusses the threat models, unique challenges, possible solutions, and significant gaps, focusing on TEE-based confidential HPC computing. We hope this discussion will improve the understanding of this new topic for HPC in the cloud and promote new research directions.
\end{abstract}

\maketitle

\chapterinitial{Confidential computing} preserves the confidentiality of data and computation while running programs on an untrusted platform, such as a public cloud. With the growing availability of high-performance computing (HPC) in the public cloud, we foresee that confidential computing will also be a need for potential HPC users who cannot access traditional HPC facilities. However, the study on the challenges and solutions for this topic is seriously lagging. 

Traditionally, HPC facilities are maintained by national labs or major research institutes and accessed by authorized users. While many industrial users \footnote{https://www.top500.org/news/why-we-care-about-industrial-hpc/} and small-institute users are potential HPC users, they may find it cumbersome to access such exclusive resources due to policies and restrictions. Since purchasing and maintaining an HPC cluster is expensive,  HPC in the public cloud is probably the most viable option for such cash-strapped users. To meet this unique demand, most major cloud providers have started offering HPC services. Researchers have done extensive studies to understand the problems with HPC in the public cloud, e.g., on viability, performance, and usability \cite{netto18}. However, no sufficient studies have been done on confidentiality issues.

In non-cloud HPC environments, the integrity of data and computing has been the primary concern in HPC security, and the HPC provider is fully trusted to guarantee the security of data and computation. Studies have been done on issues such as hardware root of trust, software and data supply chain security, and identity management. However, confidential processing of sensitive assets, including data and possibly algorithms, is a unique feature and will be an emerging demand for outsourced HPC applications. Specific examples may include but are not limited to intellectual property protection, data or algorithm embargo, and legal requirements on private data. Due to the concerns about curious or malicious insiders, co-tenants, and external attackers \cite{khoda22},  users have hesitated to move sensitive data and computation to the public cloud.

Confidential computing techniques are becoming more practical in recent years due to the Trusted Execution Environment (TEE) development. TEE creates a \emph{secure enclave} for running programs securely with the specific CPU instructions \cite{sgx-explained}. Most recent Intel, AMD, and ARM CPUs have implemented the TEE concept. Many cloud providers have started to provide TEE-enabled servers, e.g., Azure has Intel SGX-enabled servers, and Google provides AMD SEV servers. TEE essentially moves the trust on cloud service providers to the CPU manufacturers and reduces the attack surface from the entire software stack to the enclave. The hardware-enabled features have significantly improved the performance over pure software-based cryptographic approaches \cite{sharma21survey}. Typical TEE applications (without handling side-channel attacks) cost only about 1.x of non-TEE applications' \cite{akram21}. During the past few years, confidential computing has been rapidly transformed from academic research to practical applications (e.g., fortanix.com), enabling new forms of computing and sharing with reduced risk of data breaches\footnote{https://docs.microsoft.com/en-us/azure/confidential-computing/use-cases-scenarios}. However, the combination of TEE-based confidential computing and HPC in the cloud remains an insufficiently explored area. 

This paper will discuss the threat models, potential challenges, and solutions for applying TEEs to public HPC clouds. Other studies may have covered interesting topics around ``HPC in the cloud'', e.g., applying cloud computing technologies to manage a traditional HPC\footnote{https://www.hpcwire.com/2018/02/15/fluid-hpc-extreme-scale-computing-respond-meltdown-spectre/}. Our study is distinct from those, as we will focus on the fundamental issues in public HPC clouds -- users' confidentiality and ownership concerns about their data and computation. 

The remaining sections are organized as follows. Section ``Thread Modeling'' discusses threat models for confidential HPC in the public cloud. We review existing confidential computing solutions in Section ``Types of Confidential Computing Solutions''. Section ``TEE for HPC in the Public Cloud'' focuses on applying TEEs to public HPC clouds, including the challenges and solutions. Finally, we conclude the discussion.

\section{Threat Modeling} \label{sec:threat}
To discuss the possible challenges and solutions, we will need to establish a clear context for applying confidential computing for HPC in the public cloud. We will focus on unique issues that distinguish public HPC cloud from traditional HPC. 

\begin{figure}[h] 
    \centering
    \includegraphics[width=.9\linewidth]{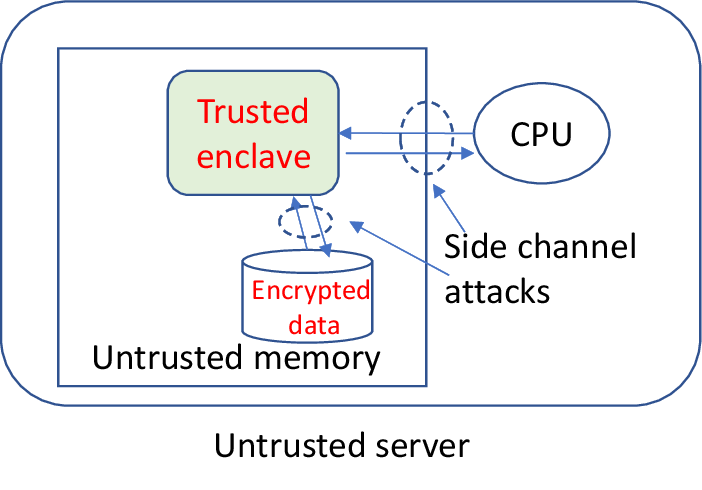}
    \caption{Threat model for TEEs.}
    \label{fig:sgx-access-pattern-attack}
\end{figure}

\textbf{Single-User Case.} Users may run confidential computation tasks in an untrusted cloud server, where the server's OS or hypervisor can be compromised. The goal is to preserve data and program integrity and confidentiality while availability is out of concern. A typical TEE, such as Intel SGX, provides a hardware-protected memory area, i.e., the \emph{enclave} \cite{sgx-explained}, and guarantees the integrity of the data and computation running inside the enclave. While adversaries cannot directly access the enclave, they can still glean information via side channels, such as memory access patterns  and CPU caches. However, cache-based attacks target all CPUs (regardless of having TEEs or not) and thus need manufacturers' micro-architecture level fixes. In contrast, the exposure of memory access patterns is inevitable as enclaves have to interact with the untrusted memory area. It's also reasonable to assume that attackers cannot access the cloud server physically, e.g., attaching a device to the server or access the motherboard, which exclude all attacks based on physical accesses.  Figure \ref{fig:sgx-access-pattern-attack} illustrates the threat model. 

\begin{figure}[h] 
    \centering
    \includegraphics[width=0.9\linewidth]{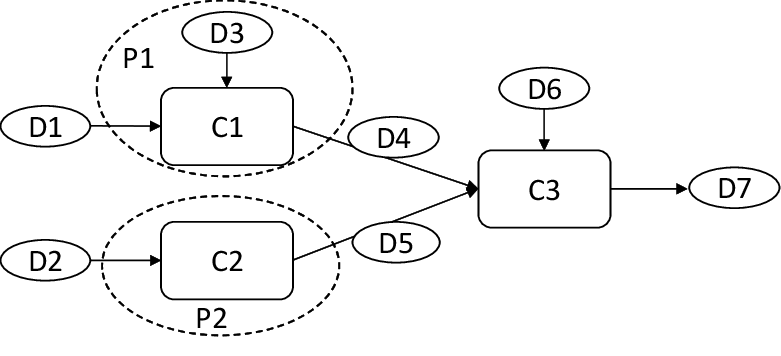}
    \caption{Collaborative workflow with private components. $P_i$: participants who may own private components, $C_i$: processing component, $D_i$: data component. $P_1$ and $P_2$ own private components, while all other components are public.}
    \label{fig:workflow}
\end{figure}

\textbf{Collaborative-Multiparty Case.} HPC applications often involve collaborative workflows, where the use of TEEs may enable new types of attacks. The following discussion also addresses general concerns with collaborative workflows, not specific to HPCs. We model a collaborative workflow as a directed graph consisting of the modules (data sources or processing modules) contributed and shared by different participants, some of which are \emph{confidential components}. Figure \ref{fig:workflow} shows two cases of confidential components in a collaborative workflow, where a participant $P_1$ holds a private dataset $D_3$ and a private algorithm $C_1$, and another participant $P_2$ has a private algorithm $C_2$ only, while all other components are public. More generally, we identify the following critical scenarios: (1) private datasets as the input or output of a processing component (either confidential or non-confidential), and (2) confidential processing components.  

Reproducibility is critical for scientific workflows. A \emph{reproducible workflow} must have a logging component to keep track of workflow provenance. Most workflow management systems, such as Galaxy and Taverna, can automatically log activities behind the scene. In contrast, users of manually built scripts and pipelines depend on public repositories, such as GitHub, to support reproducibility, using workflow scripts + a Readme file describing the inputs/outputs of each step. We consider \emph{logging}, \emph{provenance data analysis} (for debugging and optimization, etc.), and \emph{reproducibility verification} are the minimal core \emph{service components} in a reproducible workflow. These service components are likely moved to the cloud for better scalability, which can also benefit from confidential computing.  

Based on the reproducible workflow model, we aim to protect two types of assets. (1) the confidentiality and integrity of private components, and (2) the integrity of service components. Intrinsically, protecting one type of asset may interfere with the other. We consider the following potential adversaries in the collaborative environment. 
\begin{itemize}
\item \underline{Curious participants} in the workflow. While the participants' major goal is to collaboratively generate results, they might be interested in learning the private data or algorithms.        
\item \underline{Dishonest owners} of private components. They are also participants with demands on confidential processing. However, they may also take advantage of the confidential computing mechanism to disguise their fraudulent activities.
\end{itemize}
We can assume all the service components are running in a trusted environment, e.g., TEE enclaves, to make the attacking surface smaller. However, the \emph{ interplay} between the private components and service components still creates new challenges.

\section{Types of Confidential Computing Solutions}\label{sec:types}
We briefly review the available solutions of confidential computing and check whether they fit HPC applications.

\textbf{Pure Software Approaches.}  For many years, researchers have studied the software approaches to achieve confidential computing \cite{sharma21survey}. We summarize them with the following categories.

\begin{itemize}  
\item \emph{Homomorphic Encryption} allows computations with encrypted data without decryption, which is ideal for computing on untrusted platforms such as public clouds. Fully homomorphic encryption (FHE) \cite{BGV12} allows any function to be implemented on encrypted data. However, FHE's high costs in implementing multiple levels of multiplication are the primary issue, despite recent improvements in ring-based implementations \cite{BGV12}. Additive homomorphic encryption (AHE) and somewhat homomorphic encryption (SHE) methods are more efficient than FHE schemes, while allowing only a small number of homomorphic multiplications. 

\item \emph{Secure Multiparty Computation (MPC)} is another approach for multiple parties collaboratively working to evaluate a known function of their inputs while keeping the data private. Garbled Circuits (GC) and secret sharing among the main MPC methods. Recent advances such as FastGC \cite{huang11} have also significantly reduced the cost of GC. However, they are still costly, as shown in several applications \cite{sharma19,mohassel17}. Optimized secret sharing has been applied in confidential machine learning  \cite{mohassel17}, which also suffer from high communication costs. 

\item \emph{Hybrid Constructions} combines AHE, SHE, and multiparty computation primitives to minimize the overall costs of protocols. A few recent studies \cite{mohassel17,sharma19} have shown such hybrid approaches are possible for data analytics, although the costs are still much higher than plaintext approaches and TEE based solutions. 
\end{itemize}

\textbf{Trusted Execution Environment.} TEEs depend on unique CPU features to allow user-specified code and data to run inside a secure enclave that even a compromised OS or hypervisor cannot breach. It is an ideal hardware-level primitive for securely running programs on top of untrusted platforms, such as public clouds, edges, and third-party service providers. The most well-known TEE is Intel Software Guard Extensions (SGX), available in most Intel server processors, starting from the Skylake CPUs in 2015. AMD EPYC CPUs (since 2017) have also included the Secure Encrypted Virtualization (SEV) feature, which makes each protected virtual machine a secure enclave. Typically, a TEE automatically encrypts memory pages when they are not used (e.g., when swapped to the disk). The encrypted memory pages are decrypted and put in a protected memory area (e.g., the enclave page cache (EPC)) that only the owner process can access. AES encryption is used to make sure good performance and strong protection. 

\begin{figure}[h] 
    \centering
    \includegraphics[width=0.8\linewidth]{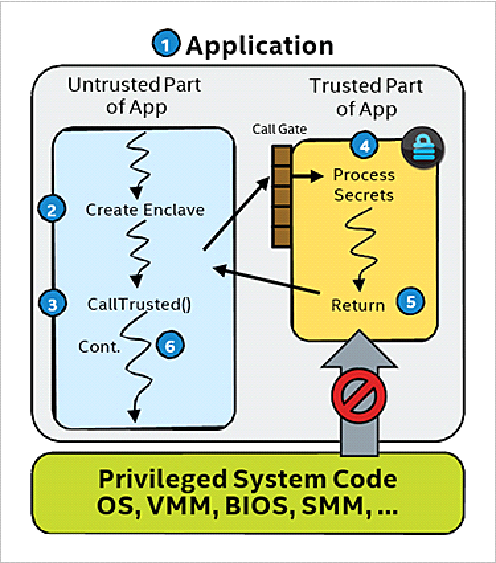}
        \caption{Illustration of SGX runtime execution (from intel.com)}
        \label{fig:sgx}
\end{figure}

Let's take a closer look at the most popular TEE implementation: Intel SGX. 
SGX implementation reserves a region of the existing system memory called Private Reserved Memory (PRM). Intel extended their x86 instruction set to isolate PRM accesses from operating systems, virtual machines, or other privilege system codes. When the user wants to perform a secure computation, it creates an isolated container known as \emph{enclave} and executes the confidential code inside the enclave. An enclave uses PRM to host data and code. Before creating an enclave, an Intel service can challenge the cloud provider via a three-party remote attestation protocol that verifies if the provider is using a certified SGX supported CPU. After creating the enclave, the user can safely upload their code to the enclave. Then, the user can pass encrypted data into the enclave, decrypt it, compute with plain text data, encrypt the result, and return it to the untrusted cloud components. During the runtime of an enclave, when other applications want to access the enclave memory, the CPU will deny such operation and return 0xFF, also known as abort page in SGX. An SGX application typically contains the untrusted part and the enclave part. Figure \ref{fig:sgx} shows SGX runtime interactions between these parts. Readers can check \cite{sgx-explained} to understand the detail.

\textbf{Discussion.} Enhanced by the hardware support, TEE programs can achieve much better performance than the pure software cryptographic approaches. The performance gain comes from three aspects. First, AES does not increase the ciphertext size and is much faster than  homomorphic encryption methods in decryption and encryption operations. 128-bit AES is typically used in TEEs. In contrast, homomorphic encryption keys are typically 1024 or 2048 bits, resulting in long ciphertext and more expensive operations. Second, TEEs have much lower communication costs than MPC solutions. TEEs only require initial remote attestation to verify the authenticity of enclaves and programs. However, MPC incurs communication costs between two cloud servers for each basic computation step (e.g., addition and multiplication operations). Finally, the computation within the enclave is done with plaintext. It's thus much faster than computation with HE or MPC. 

All these methods can preserve data confidentiality well. However, software cryptographic approaches do not protect code confidentiality, while we can also implement code confidentiality with TEEs. Both types of methods still suffer from side-channel attacks, which will be discussed in more detail later for TEEs. In contrast, protecting side channels is a lower priority in the research of pure software approaches, as other issues, such as performance and protocol-level security, have not been fully addressed yet. 

\section{TEE for HPC in the Public Cloud} \label{sec:tee-hpc}
As pure software cryptographic approaches take much more costs than the TEE approach, we believe the TEE approach is more practical for HPC applications in the cloud. There are still several challenges to using TEEs for typical HPC users. We summarize the challenges and discuss possible solutions in this section.

\subsection{Unique Challenges}
While TEEs guarantee good performance and strong security, several unique challenges exist. We summarize the main ones: usability, performance, side-channel attacks, and attacks in collaborative workflows.

\textbf{Usability}. Developing TEE applications may not be straightforward. For example, the code needs to be redesigned for Intel SGX: the application has to be split into two parts: the enclave program and the program in untrusted memory. Also, the enclave part of the code cannot use OS API directly to ensure a strong security guarantee. The learning curve will be steep for normal HPC users unfamiliar with the security concepts and the particular programming paradigm. AMD SEV does not require applications to be redesigned. However, if a higher level of confidentiality is desired, i.e., making programs resilient to side-channel attacks, the developer must modify the applications. Revising existing code is particularly unfriendly to HPC applications as most depend on low-level scientific computing libraries that have been used for decades.

\textbf{TEE Side-Channel Attacks.} Since most kinds of possible attacks in the conventional environment are no longer possible in TEEs, researchers focus on side-channel attacks. Memory side channels, i.e., access patterns, are the major ones for data-intensive processing. To be processed, encrypted data must be loaded from the untrusted areas, such as the non-TEE memory area or the file system, and then fetched by the enclave programs inside the TEE. Thus, interactions between the TEE and the untrusted area always exist regardless of what type of TEE is used, and they will be observed by adversaries and utilized to infer sensitive information. Oblivious RAM (ORAM) \cite{sasy18} has been a popular method to hide block-level access patterns for TEEs, such as ZeroTrace \cite{sasy18}, Obliviate, and Oblix.  Researchers have also used page-fault interrupts, and page table features to extract secrets such as encryption keys inside enclave \cite{xu15}. The most popular method to address this problem uses the CMOV instruction to rewrite each branching statement \cite{olga16,sasy18,alam21} to make them oblivious. The CPU cache is another popular side channel. Attacks like Meltdown and Spectra \cite{kocher19} apply to both Intel and AMD CPUs, and TEEs are not immune to such attacks. However, the defenses against cache-related attacks often depend on manufacturers' micro-architecture level firmware or software fixes.

\textbf{Performance.} In general, TEE will have a performance penalty depending on types of workloads and CPUs. Akram et al. \cite{akram21} have shown that with proper configurations the cost for HPC benchmarks can be around x1.15 slowdown for AMD SEV, while varying in a larger range for Intel SGX. Note that these tests do not consider any side-channel protection mechanism. The current access-pattern protection mechanisms, such as ORAM will significantly increase the overhead -- reducing the cost of protection has been one of the primary goals for ongoing research \cite{alam21}.

\textbf{Issues with Collaborative Workflow.}
As HPC applications typically involve workflows, some of which may also include multiple parties, confidential computing in this context also raises new problems. 
\begin{itemize}
\item \emph{Owner's Attacks.} Private components may create a blind spot in the workflow system: They do not allow other users to examine the internal details, and the logging service only records the external information about using the algorithm component, i.e., parameter settings, input, and output. Thus, dishonest private-component owners have a chance to issue a \emph{replacement attack} that the owner can replace the private algorithm or data anytime. This attack is even more challenging to detect for algorithms with randomization steps, which are common in scientific computing. A dishonest owner can forge a fake algorithm that works only for specific input-output pairs while replacing it later with another one that generates outputs with a similar statistical property. 

\item \emph{Conflict between Confidentiality and Provenance Analysis.} Provenance analysis needs to access the log data, which includes the description of the activities around a private component (a dataset or a processing program).  As a result, attackers may utilize this information to infer the content of the private component. For example, model-inversion attacks \cite{fredrikson15} are possibly applied to infer private input data from a known processing algorithm and output data; and model-stealing attacks \cite{tramer16} try to rebuild the private processing algorithm based on sample input-output pairs.

\item \emph{Reproducibility Verification} is a replay of workflow execution, supposedly conducted by an authorized third party. Due to the security requirements (e.g., passing secret keys to the enclave), TEE-based private components are controlled and executed by their owners, which have effectively prevented any unauthorized verification. However, it's inconvenient to demand all owners staying online for verification. Another concern is that the dishonest owner's attack can also be applied in this stage.  
\end{itemize}

\subsection{Possible Solutions}
The application of the TEE approach is still at the early stage. In the following, we discuss some solutions that can be applied to HPC applications in the cloud.

\textbf{Improving Usability.} 
A few efforts have been conducted to usability issues of SGX. To avoid modifying existing applications, Graphene-SGX \cite{tsai17} and SCONE \cite{arnautov16} try to build a library OS or a shim layer to allow unmodified Linux applications running inside enclaves. However, this approach does not protect access patterns from attacks, and it's difficult to incorporate any application-level protection methods into these frameworks. Other approaches, such as Google Asylo and Open Enclave, try to simplify SGX programming with an easier programming framework or library, so that users do not need to learn the complex native SGX APIs.  

For distributed data-intensive processing, VC3, M2R, and Opaque \cite{zheng17} try to adapt existing popular data-processing software stacks such as Hadoop and Spark by slightly modifying the original software, e.g., only moving the confidential data processing part into the enclave. However, leaving the system components in the untrusted memory area enables many attacks. 

\emph{\underline{Gaps:} To our best knowledge, all these usability-oriented projects do not address the side-channel attacks. However, these methods are particularly useful for legacy HPC applications, if side-channel protection is not a concern.      
}

\textbf{Protecting Block Access Patterns with ORAM.} The ORAM-based approaches \cite{sasy18} try to disguise block I/O accesses and provide a generic block I/O interface for applications. However, ORAM-based methods have several notable drawbacks. (1) They are expensive. Each block access incurs $O(\log n)$ additional block accesses to disguise the actual access, where $n$ is the number of blocks in the file. (2) Not all block access patterns leak sensitive information, which requires users to examine the application-specific block access patterns to design solutions with good performance, which is time-consuming and error-prone. (3) As a low-level I/O interface, they do not aim to protect application-level access pattern problems such as branch prediction attacks. 

\textbf{Application-Specific Data Oblivious Approaches.} The application-specific approach requires developers to carefully examine all access patterns of a specific application and apply \emph{oblivious operations} to protect them. Ohrimenko et al. \cite{olga16} have analyzed a batch of well-known machine learning algorithms and identified that oblivious move (omove), oblivious greater(ogreater), and oblivious sorting (osort) are the three most frequently needed operations by these algorithms. omove and ogreater use the CMOV instructions to achieve obliviousness, which have been mentioned earlier. For experienced developers, this approach can thoroughly address all access-pattern problems. However, again, it is time-consuming and probably not practical for most HPC developers.  

\begin{figure}[h]
\centering
\includegraphics[width=0.9\linewidth]{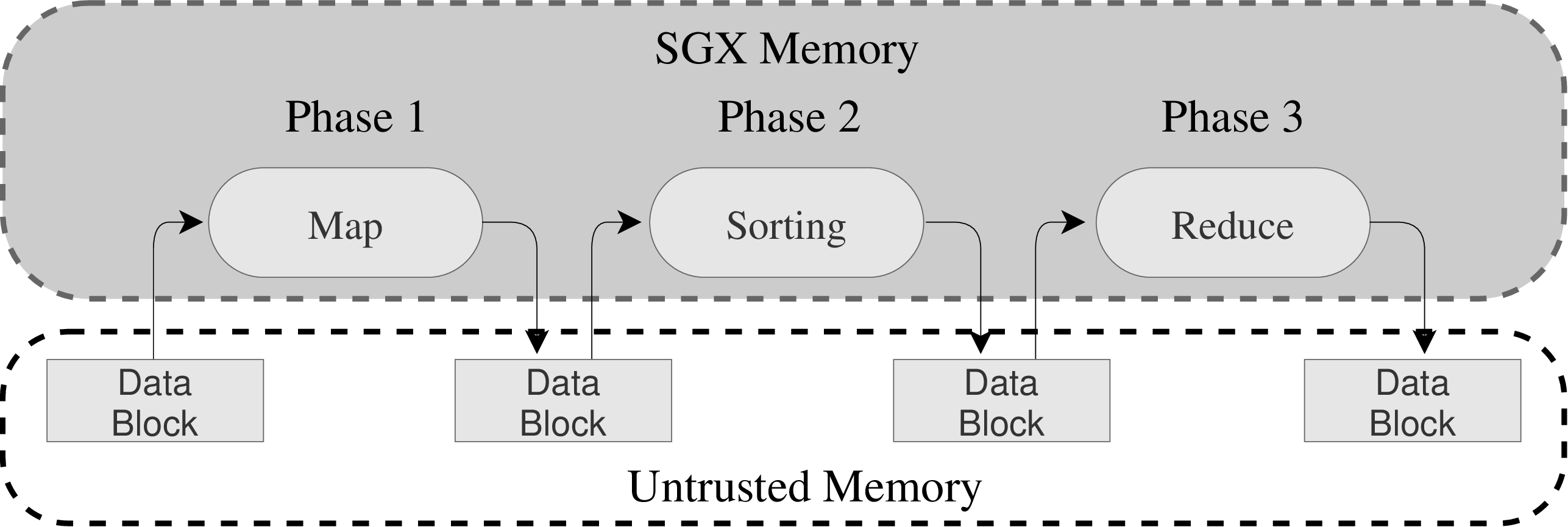}
\caption{Regulated data flow between enclave and main memory (from SGX-MR \cite{alam21})}
\label{fig:regulated-dataflow}
\end{figure}

\textbf{Framework-Level Access Pattern Protection.} A more promising solution is the \emph{framework-level} protection scheme, such as SGX-MR \cite{alam21}, which achieve a balance between usability and access-pattern protection. The idea of SGX-MR is to regulate the application data flow with a framework, and then identify and protect the access patterns of the regulated data flow and the within-block (or page) access patterns. Once these access pattern problems are addressed, all applications using this framework will benefit. MapReduce is the right candidate for this purpose. Figure \ref{fig:regulated-dataflow} shows how  the application data flow is regulated by the MapReduce processing pipeline at the block level. (1) The input to the Map phase is just sequential reads, not leaking any information. (2) The Sorting phase can use an efficient \emph{oblivious sorting} algorithm. (3) In the Reduce phase, we can protect the output privacy, i.e., group sizes. Users only need to implement the \texttt{map}, \texttt{reduce}, and possibly \texttt{combine}, functions, which only need to handle in-page branching statements -- these are typically much easier to handle.

SGX-MR has unique advantages in transparency, programmability, efficiency, and attack protection. (1) \emph{Transparency} is achieved with a carefully designed middle layer between TEE and user applications. It hides all the details about TEE processing and access-pattern protection.  (2) It provides reasonably good \emph{programmability}. Instead of emulating OS APIs, this approach utilizes the MapReduce processing framework to unify the applications at the framework level. Almost all existing data mining and machine learning algorithms can be implemented with one or multiple MapReduce programs, as shown by Mahout and numerous examples during the past ten years with the booming big data applications.  (3) This design can achieve better \emph{efficiency} in access pattern protection than ORAM-based approaches \cite{sasy18}, while the difficulty of redesigning users' applications is much lower than the customized data-oblivious approaches \cite{olga16}.  (4) The framework also allows users to achieve different levels of protection against application-oriented and memory page-oriented access-pattern attacks to meet various demands on performance and usability. 

The goal of SGX-MR is not to provide a framework for legacy applications, as rewriting the code to use SGX-MR is not easy for most applications. It's more appropriate for those data-intensive applications that can be easily modified or developed from scratch.  In addition, SGX-MR is still at the preliminary stage focusing on single-node processing. In contrast, multi-node processing is the norm for HPC applications. Opaque \cite{zheng17} has mentioned several access pattern issues with inter-nodes data exchange, which should be integrated into the extension of SGX-MR to multiple nodes.

\emph{ \underline{Gaps:} The above three approaches address the access pattern problem. (1) While new HPC applications can use these attack-mitigation methods, to our best knowledge, there is no solution to address side-channel attack problems for legacy HPC applications yet. In particular, as many HPC applications depend on scientific computing libraries, making these libraries fully data oblivious is very challenging. (2) ORAM and SGX-MR target data-intensive applications, but many HPC applications are compute-intensive, where new framework-level access-pattern protection methods should  be developed. (3) All these data oblivious methods will impair performance, the level of which has not been fully understood yet for HPC applications. }

\textbf{Monitoring and Detecting Side-channel Attacks.} This approach is attractive as it may avoid revising the existing codes (e.g., after using Graphene-SGX or SCONE to achieve good usability). The idea is to monitor the abnormal patterns of page fault interrupts or other system-level activities to detect possible attacks, as many attacks utilize these system features. While an attack is in progress, these system-level activities might differ from normal program execution. For example, SGX-TSX \cite{shih17} has followed this approach. It utilizes Intel Transactional Synchronization Extensions (TSX) to monitor page-fault interrupts. TSX is a CPU built-in mechanism and cannot be compromised by attackers. Using TSX transactions, SGX-TSX detects anomalies and terminates the enclave programs as needed.

\emph{ \underline{Gaps:} The monitoring and detection approach is promising for protecting legacy code that cannot be easily modified. However, the current method: SGX-TSX is not easy to use and still requires a certain level of code modification. Another issue is false alarms, which might accidentally interrupt normal programs.}

\textbf{Blockchain-based Workflow Management.} As we have discussed, collaborative HPC workflows may bring more challenges: dishonest owners, the conflict between confidentiality and provenance analysis, and the inconvenience in reproducibility verification. We envision a blockchain-based solution that can probably address most of these problems. 
\begin{itemize} 
    \item \emph{Protect from dishonest owners.} Use the blockchain to store the non-fungible signatures of the program and data. While this does not prohibit users from uploading fake data or algorithms or tampering with data and algorithms, we can trace the exact version used in a specific run. 

 \item \emph{Control accesses to provenance data.}  Control the access to provenance analysis and log the accesses for anomaly detection. Access control can be reinforced with blockchain-maintained logs and smart contracts. We can also build an anomaly detection subsystem, learning from the tamper-resistant provenance access log. The challenge is to develop an effective anomaly detection algorithm using the provenance access patterns. We can also prohibit access to the provenance data related to private components or their nearby components, which will significantly reduce the utility of provenance data.  

 \item \emph{Automated secure replay of workflows} can be implemented with smart contracts, which do not need owners of private components to stay online. It also prevents any attacks trying to compromise the integrity of reproducibility verification.
\end{itemize}

\emph{ \underline{Gaps:} (1) Blockchain applications are still in the embryonic stage. The cost of using current public blockchains is too high to be practical, while permissioned blockchains, such as HyperLedger, will need users to trust the management peers. (2) The access control and anomaly detection methods for provenance analysis will deter some attackers, but do not eliminate the risk of attack -- they merely increase the chance of attackers getting caught. 
}

\section{Conclusion}\label{sec:conclusion}
HPC in the cloud can benefit many users who cannot own or access on-premise HPC resources. Recent studies have explored several aspects of HPC in the cloud, while the confidentiality issues have not been addressed yet. As data and computation confidentiality has been a general concern for many cloud users, we anticipate that HPC users may also have such needs in the future. Confidential computing has become practical due to the recent development of trusted execution environments, but it is still at the early stage of applications. We envision that combining TEE and HPC may raise some unique challenges, especially in a collaborative environment. We have analyzed the threat models for the single-user and collaborative-workflow cases, discussed several unique challenges, including usability, side-channel attacks, performance, and the interplay between confidential components and collaborative workflow, and reviewed some candidate solutions. We have also highlighted a few gaps that appear no satisfactory solutions yet, which probably indicate valuable research directions. 
\section{Acknowledgements}
This work was supported in part by the U.S. National Science Foundation under Grant \#2232824, Marquette University, and Northwestern Mutual Data Science Institute. 

\bibliographystyle{IEEEtran}
\bibliography{./references.bib}

\end{document}